\documentclass[twocolumn, prl]{revtex4}
\usepackage{amsmath}
\usepackage{amssymb}
\usepackage{graphicx}
\usepackage{bbm}

\begin{document}
\title{Knotted Nematics}
\author{Thomas Machon}
\author{Gareth P. Alexander}
\affiliation{Department of Physics and Centre for Complexity Science, University of Warwick, CV4 7AL, UK}

\date{\today}

\begin{abstract}
Knotted line defects in continuous fields entrain a complex arrangement of the material surrounding them. 
Recent experimental realisations in optics, fluids and nematic liquid crystals make it important to fully characterise these textures and to understand how their properties relate to the knot type. 
We characterise knotted nematics through an application of classical knot theory founded upon the Pontryagin-Thom construction for nematic textures and give explicit closed form constructions for knots possessing Milnor fibrations with general boundary conditions. For links we construct nematic textures corresponding to all possible assigments of linking numbers and discuss the relevance to recent, and classic, experiments on Hopf links. 
\end{abstract}
\maketitle

Knots have fascinated for millennia, inspiring ancient artwork and the beautiful illustrations of the Book of Kells~\cite{kells,przytycki98,jablan12}. In science, they appear naturally in DNA, can be synthesised into chemical topology, and continue to motivate the latest developments in mathematics~\cite{jones85,khovanov00} and physics~\cite{witten89,kitaev03,dennis10,kleckner13}. Surprisingly, the first serious mathematical study of knots was not until Tait's tabulation, in the late 19$^{\text{th}}$ century, of the first seven degrees of knottiness~\cite{tait76}, inspired by Kelvin's ill-fated `vortex atom' theory~\cite{thomson67}. Tait's knots were familiar line drawings of knotted curves, whereas Kelvin's vortices were knots in a continuous field. In a field there is more than just the knotted curve; the entire material is imbued with a complex structure, examples of which have been studied in fluid vortices~\cite{kleckner13,moffatt69,liu12}, electromagnetism~\cite{ranada92,irvine08}, particle physics~\cite{brekke92,faddeev97,sutcliffe07}, and quantum computing~\cite{kitaev03,nayak08}. Materials with non-trivial fundamental group support line defects, which can be tied into knots; like Kelvin's vortices these are inseparably embedded in a continuous field and their full characterisation involves the complex {\em global} structures described here. In liquid crystals, such knotted fields have only recently been realised experimentally~\cite{tkalec11,jampani11}, reinvigorating interest in understanding how their properties relate to those of the knot and how they might be controlled~\cite{machon13}. 
Here, using the Pontryagin-Thom construction~\cite{dieck08,chen13,BryanThesis}, we characterise an array of knotted nematics through the tools of classical knot theory, and accompanying this present an explicit method of construction for a broad class of experimentally realisable knotted textures. For simple linked loops we identify a distinction between textures that has not been considered in experimental realisations, but is readily accessible. This construction provides a framework for the analysis of knotted nematic textures, experimentally, theoretically and computationally.

Nematic liquid crystals support both point defects (hedgehogs) and line defects (disclinations) through the first and second homotopy groups of the ground state manifold (GSM), which is the real projective plane $\mathbb{RP}^2$, corresponding to a unit line field. The homotopy groups  $\pi_1(\mathbb{RP}^2)\cong\mathbb{Z}_2,\,\pi_2(\mathbb{RP}^2)\cong\mathbb{Z}$, which interact through the action of $\pi_1$ on $\pi_2$~\cite{volovik77,mermin79}, describe the possible local behaviour of both point and line defects. For line defects the fundamental group ($\pi_1$) measures the topology on a circle going around the disclination. More information is given by the configuration on a torus enclosing the entire loop~\cite{bechluft-sachs99}, which yields an index $\nu \in \mathbb{Z}_4$~\cite{alexander12,janich87} in the case of $\mathbb{RP}^2$. While these methods yield robust results, their classifications are based on {\em local} information: the defect line itself and the texture in a small neighbourhood. However, to understand the full possibilities of these configurations it is necessary to view the {\em global} information. The Pontryagin-Thom (PT) construction~\cite{dieck08} supplies this global information. When applied to nematics~\cite{chen13,BryanThesis}, which are modelled using unit line fields ${\bf n}({\bf r})\sim -{\bf n}({\bf r})$ called the director, it reduces ${\bf n}({\bf r})$ down to its essential topological data, from which the entire texture can be reconstructed up to smooth deformation. One chooses a particular point ${\bf p}$ in the GSM ($\mathbb{RP}^2$) and then constructs the surface where the director is perpendicular to ${\bf p}$. The boundaries of this surface are the defects in the system. There is an additional degree of freedom corresponding to the director orientation in the plane perpendicular to ${\bf p}$ and this is used to colour the surface. 
This construction is essential for gaining an intuitive understanding of knotted textures and in general is an excellent tool for identifying the topological content of ordered media. 

\begin{figure}[t!]
\includegraphics{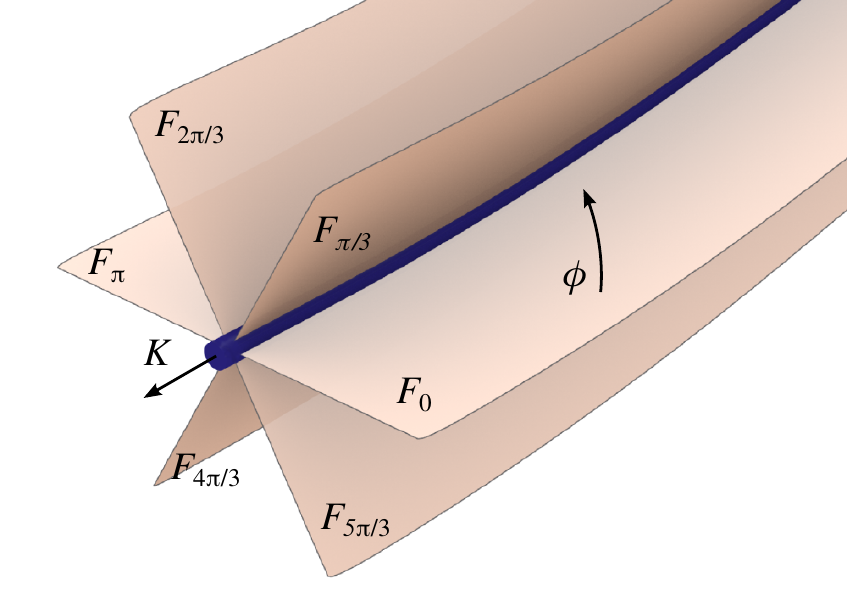}
\caption{(Colour Online) Structure created by Milnor fibrations. The knot, $K$, is the common boundary to all the fibres, $F_\phi$. The fibres themselves are indexed by $\phi=\textrm{arg}(f)$ which itself rotates through $2\pi$ as one makes a small loop around $K$. The direction of increasing $\phi$ induces an orientation on $K$.}
\label{fig:1}
\end{figure}

To generate knotted configurations we exploit mathematical structures known as Milnor fibrations, previously applied in optics and electromagnetism~\cite{dennis10,kedia13,arrayas13}. These structures, described in Milnor's seminal 1968 book~\cite{milnor68} and a rich subject in their own right~\cite{seade05}, allow one to relate many topologically interesting objects to the singular points of simple complex (and real) polynomials. The singular points of such a polynomial, $f: \mathbb{C}^n \to \mathbb{C}$, with an isolated critical point at the origin are studied by restricting $f$ to $S^{2n-1} \subset \mathbb{C}^n$, {\it i.e.} $\sum_i |z_i|^2 =1$. The zero set of $f$, denoted by $K$, can correspond to a variety of mathematical objects; we will consider $n=2$, for which one obtains knots and links~\cite{note1}. Defining a new function $\phi=\textrm{arg}(f)$, we find that $K$ becomes a singularity of $\phi$, around which $\phi$ winds through a full $2\pi$, as shown in Fig. \ref{fig:1}.  If this is the case then $\phi$ defines a fibration of $S^3-K$. This means that locally (but not globally) the space looks like $I \times F_\phi$ where $I \subset \mathbb{R}$ is an interval and $F_\phi$ is a set of constant $\phi$, called a fibre. These fibres, indexed by $\phi$, form oriented surfaces whose boundary is the knot or link, \textit{i.e.} they are Seifert surfaces of $K$~\cite{rolfsen03}.

The simplest class of knots that can be created in this way are the torus knots, constructed using Pham-Brieskorn polynomials~\cite{pham65,brieskorn66} $f(z_1,z_2)=z_1^p+z_2^q$ where $p,q$ are positive integers. These polynomials have a zero set, $K$, equal to a $(p,q)$ torus knot or link~\cite{brauner28} (it is a knot only if $p$ and $q$ are co-prime).

\begin{figure}
\includegraphics{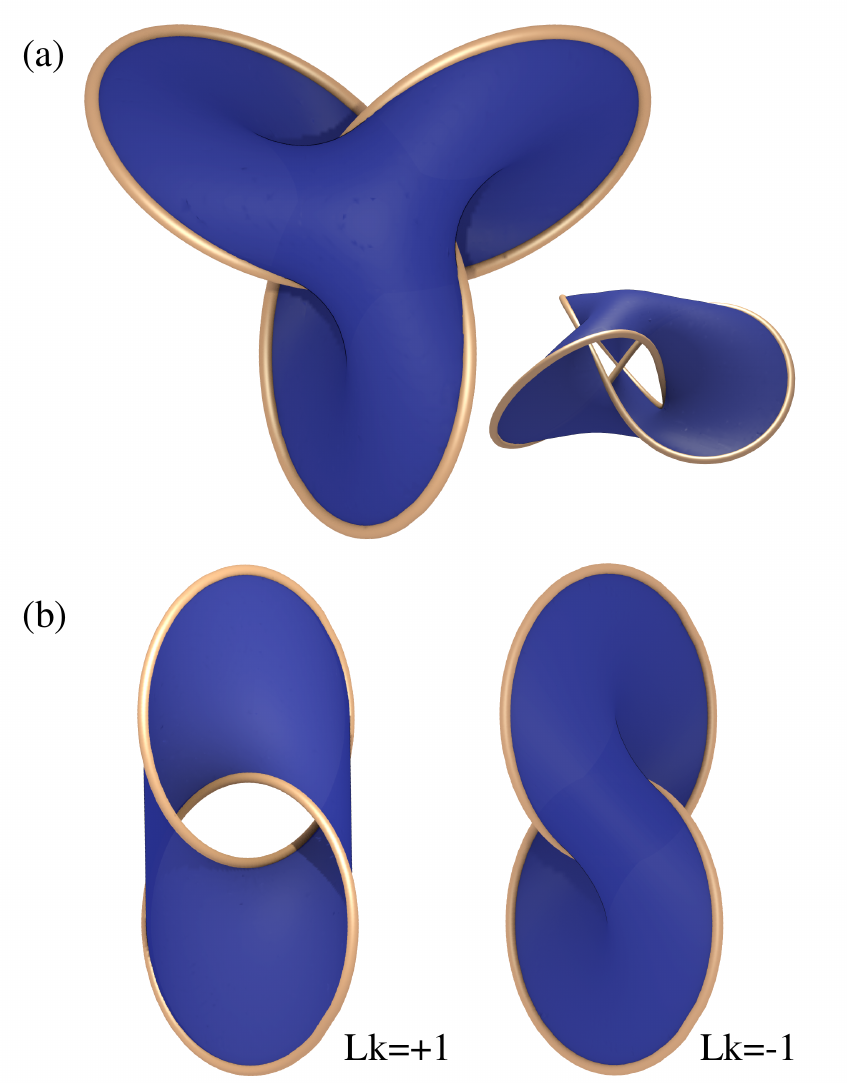}
\caption{(Colour Online) PT surfaces for knotted disclination lines in planar textures. The PT surface consists of all points where the director lies in the $x$-$y$ plane, and is a Seifert surface for the knot or link in question. Due to the the structure of $\textbf{n}$ in this case, the surface is simply the stereographic projection into $\mathbb{R}^3$ of the fibre $F_\pi$, on which the director points in a constant direction, along the $x$ axis. Since the director does not rotate in the $x$-$y$ plane, the surfaces are coloured uniformly. (a) Trefoil knot obtained with $p=3$, $q=2$. The PT surface here is of genus 1, the genus of the trefoil knot. (b) Distinct Hopf link textures with linking numbers $\text{Lk}=+1$ and $\text{Lk}=-1$.}
\label{fig:2}
\end{figure}

To move this structure into $\mathbb{R}^3$ we employ stereographic projection, which maps all of $S^3$ into $\mathbb{R}^3$ except a single projection point, which we take to be $(0,i)$. $\phi$ tends to a constant at large distances from the $\mathbb{R}^3$ origin; to control this we use the modified set of polynomials $f(z_1,z_2)=z_1^p+(-iz_2)^q$, which ensures that $\phi\to 0$ at large distances.

This construction gives a function containing a knotted singularity, $\phi:\mathbb{R}^3-K \to S^1$. If the GSM of our system is itself $S^1$ (as in optics \cite{dennis10}) then we are done, but for more general knotted fields a map $S^1\to \text{GSM}$ sends the phase of the Milnor map to the ground state manifold, specifying the type of the knotted defect through the element of $\pi_1(\text{GSM})$ represented by the map. For nematics, the GSM is $\mathbb{RP}^2$ and $\phi$ embeds as a non-trivial cycle. For example, choosing the director to lie in the $x$-$z$ plane we may write
\begin{equation}
\textbf{n}(\textbf{r})=\left(\sin \left(\frac{\phi(\textbf{r})}{2} \right),0,\cos \left(\frac{\phi(\textbf{r})}{2} \right)\right),
\label{eq:c0}
\end{equation}
which defines a knotted disclination. Taking a small loop around a cross-section of $K$ will give a winding of $2\pi$ in $\phi$, sending $\textbf{n} \to -\textbf{n}$ and creating the familiar $\pi$ winding required for a disclination~\cite{note5}. By construction this texture is planar, with $\textbf{n}=\hat{{\bf z}}$ as $r \to \infty$ and the PT surface is just the projection of the fibre $F_\pi$ on which $\textbf{n}$ is constant, inducing a uniform colouring, shown in Fig. \ref{fig:2}.

These constructions via Milnor fibrations always generate an orientable PT surface. This orientability induces a relative orientation on the boundary components so that the Hopf link texture, for example, acquires a well-defined linking number, shown in Fig.~\ref{fig:2}(b). The polynomial for the Hopf link, $(z_1+z_2)(z_1-z_2)$, is a product, with each factor corresponding to a component of the link. The construction of \eqref{eq:c0} gives a nematic texture with linking number $\text{Lk}=+1$. A texture with linking number $-1$ can be constructed by using instead the polynomial $(z_1+z_2)\overline{(z_1-z_2)}$, that differs only in taking the complex conjugate of one of the factors. 

Linked loops were first observed by Bouligand in 1974~\cite{bouligand74} and have recently been made around colloids~\cite{jampani11} but the full nature of the nematic texture has not been studied, something that is readily available with current techniques~\cite{chen13}. While in an experiment the actual texture will not necessarily be close to the perfect planar textures of \eqref{eq:c0}, (indeed the texture may even have a non-orientable PT surface) in the case of the Hopf link it will always be homotopic to one of the two planar textures shown in Fig.~\ref{fig:2}(b) \cite{machon13b}. 

This construction for textures with different linking numbers naturally extends to $n$-component links arising from polynomials of the form $f=f_1 f_2 \ldots f_n$. Knotted textures corresponding to each of the $2^{n-1}$ different combinations of linking numbers can be constructed by conjugating appropriate factors in the polynomial. 

\begin{figure}[t]
\includegraphics{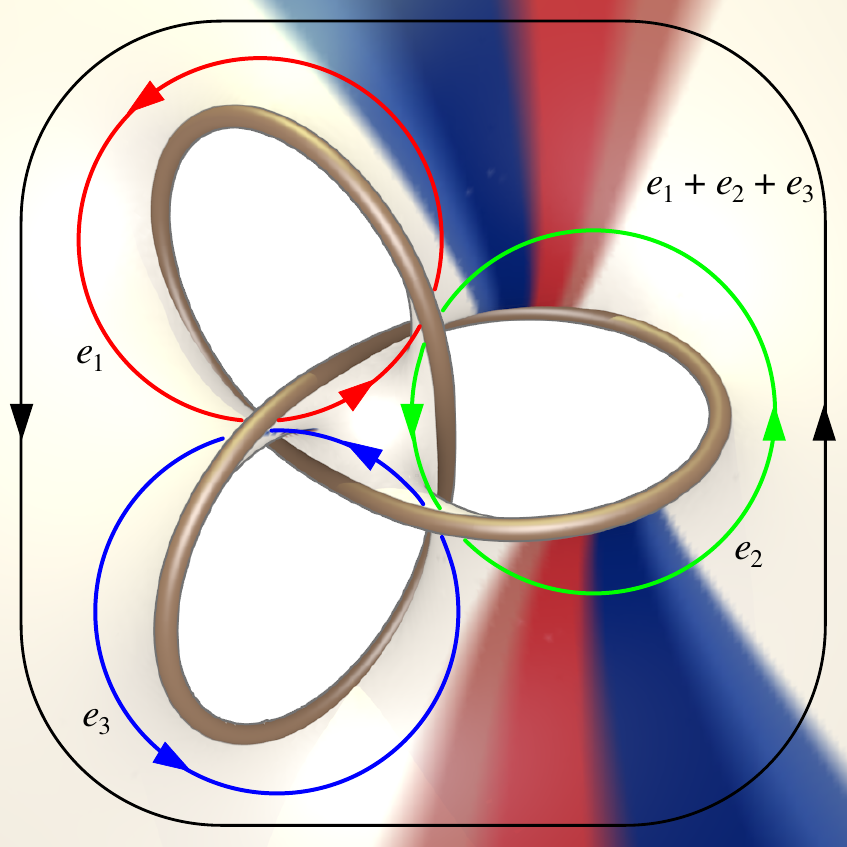}
\caption{(Colour Online) Example PT surface, $F$, for a charge one trefoil knot. The surface can be viewed as a sheet that stretches out in the $x$-$y$ plane, connected to the small central region by three twisted strips. The colour winding shows the orientation in the $x$-$y$ plane, and is artificially contracted to emphasise the topology. A loop traced around the knot at large distances goes around the colour wheel twice, identifying the knot as having unit charge. The three cycles, $e_1$, $e_2$ and $e_3$ form a basis for the first homology of $F$ on which colour windings can be distributed, with the total charge being the winding on the cycle $e_Q=e_1+e_2+e_3$. In this case there is a double winding on $e_2$, giving a total charge of one.}
\label{fig:3}
\end{figure}

Equation \eqref{eq:c0} only produces planar textures with uniform director at large distances. Experimentally, however, nematics are often confined to different geometries with non-uniform boundary conditions, such as droplets where the boundary conditions may be radial~\cite{poulin97}, or even droplets with handles~\cite{pairam13}. Such systems can support knotted defects~\cite{sec13}, but the boundary conditions in these environments mean that the textures cannot be planar, so the use of Milnor fibrations must be adapted to construct representatives in these classes. 

\begin{figure*}[t]
\includegraphics{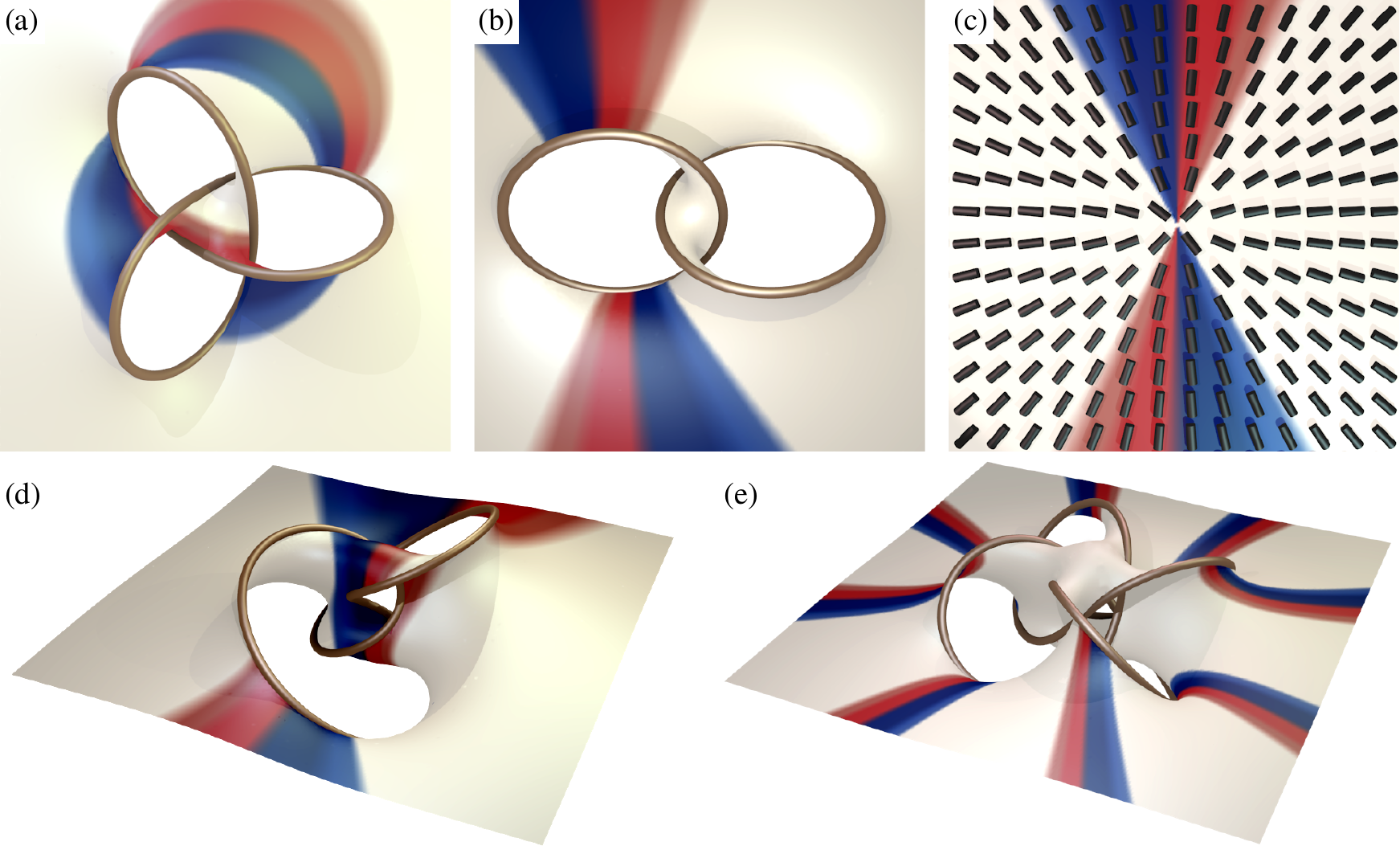}
\caption{(Colour Online) Knotted nematics and associated PT surfaces constructed using \eqref{eq:2} and \eqref{eq:3}. (a) $p=3$, $q=2$ trefoil knot, with colour windings of -2, 2 and 0 on the three basis cycles $e_1, e_2, e_3$. The total charge is zero, implied by the colour being uniform at large distances. (b) $p=2$, $q=2$ Hopf link, with two-fold winding on one cycle, showing the localisation of the hedgehog charge around one of the disclinations. (c) Legend showing the director-colour correspondence. The colour winding is deliberately contracted to emphasise the topology of the knots and links. (d) $p=2$, $q=3$ trefoil knot, the $(p,q)$ and $(q,p)$ torus knots are equivalent. This knot has a global two-fold winding, giving it charge one. (e) $p=4$, $q=3$ torus knot with four $+2$ windings, giving a total hedgehog charge of 4.}
\label{fig:4}
\end{figure*}

A generalisation of \eqref{eq:c0} which generates such textures is given by 
\begin{gather}
\textbf{n}= \cos \left( \frac{\phi}{2} \right ) \textbf{h} - \sin \left( \frac{\phi}{2} \right ) \hat{{\bf z}} ,
\label{eq:2} \\
\textbf{h}= \bigl( \sin(\chi)\cos(\theta),\sin(\chi)\sin(\theta),\cos(\chi) \bigr) ,
\label{eq:3}
\end{gather}
where $\chi$ is the usual polar angle, $\theta = \Im\, \textrm{log}\, \Theta$ and $\Theta (x+iy,z) : \mathbb{C} \times \mathbb{R} \to \mathbb{C}$ is a smooth one-parameter family of meromorphic functions of $x+iy$, indexed by $z$. Examples are shown in Figs.~\ref{fig:3},~\ref{fig:4} as we now explain. The Milnor phase $\phi$ rotates the director between being vertical and lying horizontally, while the angle $\theta$ controls the horizontal orientation. The PT surface $n_z=0$ given by this method is an algebraic surface defined implicitly by $z/r-\tan(\phi/2) \equiv z/r-\Im (f)/(\Re (f) + |f|)=0$ that tends to the $x$-$y$ plane at large distances. Since it is independent of $\theta$, the horizontal orientation of the director can be varied without altering the PT surface. On a large sphere enclosing the knot the texture induces a homotopy class $Q\in\pi_2(\mathbb{RP}^2)$. Such behaviour can match with no further defects onto the surface of a droplet with $\gamma$ handles provided $Q=1-\gamma$~\cite{poulin97,pairam13,stein79} and is ensured by the asymptotic behaviour $\Theta\sim(x+iy)^Q$. In terms of the colouring of the PT surface, this corresponds to a $2Q$-fold colour winding far from the knot. 

How does this large distance colour winding connect onto the knot? The PT surface from \eqref{eq:2},~\eqref{eq:3} is orientable so it provides a Seifert surface, $F$, for the knot~\cite{rolfsen03}. A Seifert surface for a given knot has a genus, $g$, and while different surfaces may have different genera, the minimum $g$ over all possible surfaces, is a knot invariant (equal to $((p-1)(q-1)+1-\textrm{gcd}(p,q))/2$ for torus knots). Such a surface with genus $g$ has $2g+b$ linearly independent cycles, where $b$ is the number of components in the link~\cite{note4}, on which the asymptotic colour winding is distributed. Any distribution of windings can be generated through the construction \eqref{eq:2}, \eqref{eq:3}. The colouring is the angle $\theta$ of the director on $F$, which winds around the poles and zeros of $\Theta$ so that the location of the poles and zeros controls the location of the colour winding on $F$. As $z$ varies the poles and zeros of $\Theta$ trace out a family of trajectories ${\cal T}$, which can be chosen to produce the desired colour winding. 
 
A distribution of colour windings over the homology cycles of $F$ corresponds to an element of the cohomology $H^1(F)\cong\mathbb{Z}^{2g+b}$. By Alexander duality~\cite{hatcher02} $H^1(F)\cong H_1(S^3-F)$ elements of the cohomology can be thought of as cycles in the complement $S^3-F$. A choice of $\Theta$ whose poles and zeros trace out any such trajectory then generates a texture with any desired distribution of colour winding. The winding on any cycle on $F$ is given by the sum of the orders of the zeros minus that of the poles of $\mathcal{T}$ that pass through a disc bounded by the cycle. If the trajectories pierce the PT surface the intersection corresponds to the location of an additional point defect in the liquid crystal; this effectively creates a puncture in the surface -- the director is not defined at the defect location -- around which an additional cycle is created. 

An example PT surface for the trefoil knot is shown in Fig. \ref{fig:3}, where we have three homology cycles, $e_1$, $e_2$ and $e_3$ on which we may distribute colour windings. The configuration shown has a single winding on $e_2$ which corresponds to a function $\Theta(x+iy,z)=x+iy -l_2$ where $l_2 \in \mathbb{C}$ is the `centre of the hole' in the PT surface that is enclosed by $e_2$. The total charge $Q$ of the knot is given by the winding on the cycle $e_1+e_2+e_3=1$. A natural construction for this ``total charge'' cycle can be given using the Seifert matrix associated to the basis $\{e_i\}$. The charge cycle $e_Q$ corresponds to the kernel of the Seifert matrix. Since the kernel is independent of any choice of basis, this gives a basis independent means of identifying $e_Q$, and hence $Q$. 

This construction allows many exotic objects to be created. For example, Fig. \ref{fig:4}(a) shows a trefoil knot defect with windings of opposite sense on two cycles and while the total charge of this configuration is zero, it carries colour winding. In terms of the function $\Theta$, a pole of $\Theta$ passes through the hole corresponding to the cycle $e_1$ and a zero through the hole corresponding to $e_2$, giving a colour winding of $-2$ on the first cycle and $+2$ on the second. Fig. \ref{fig:4}(b) shows a Hopf link of charge $+1$. The colour winding can be associated with the left of the two components, identifying it as the charged component. $\Theta$ has a single zero along a trajectory passing through the charged component. 

We conclude with some caveats and speculation. Firstly, the textures constructed here only generate the specified topology, they are not necessarily minimisers of any physically relevant free energy although they may be metastable. Secondly, the full topological equivalence of these textures is still unknown. While for a fixed PT surface the configurations described here are all distinct, these surfaces can be altered by certain bordism moves that that are consistent with maps from the normal bundle of the surface into $\mathbb{RP}^2$, and the full zoo of knotted textures described here is likely an over-complete list. Indeed, such analysis for the Hopf link reduces the number of charge zero textures to just two -- the left and right handed links of Fig.~\ref{fig:2} -- something that will be explored in future work~\cite{machon13b}. Finally, it would be of great interest to analyse knots produced experimentally in the way presented here.

We are grateful to Ryan Budney, Bryan Chen, Mark Dennis, Randy Kamien and Miha Ravnik for beneficial discussions. This work was partially supported by the EPSRC. TM partially supported by a University of Warwick Chancellor's Scholarship. GPA acknowledges partial support by NSF Grant No. PHY11-25915 under the 2012 KITP miniprogramme ``Knotted Fields''.


\begin{thebibliography}{10}
\bibitem{note5} While in principle we could consider any odd $\pi$ winding, higher windings are rarely observed experimentally and we will not discuss them here.
\bibitem{note4} One would usually expect $\textrm{dim}(H_1(F)) =2g-1+b$. The extra $+1$ in the dimension comes from the additional hedgehog at infinity, which creates a once punctured Seifert surface for the knot or link. We also mention that this is the minimal number corresponding to where $g$ is knot genus rather than just the genus of the surface.
\end{thebibliography}
\end{document}